\documentstyle[twoside,fleqn,epsf,espcrc2]{article}

\newcommand{\AmS}{{\protect\the\textfont2
  A\kern-.1667em\lower.5ex\hbox{M}\kern-.125emS}}
\newcommand{\ewxy}[2]{\setlength{\epsfxsize}{#2}\epsfbox[10 60 640 570]{#1}}

\title{The static potential beyond screening in the 3d SU(2) Higgs model} 

\author{O. Philipsen\address{Institut f\"ur Theoretische Physik, 
        Philosophenweg~16, D-69120 Heidelberg, Germany}
        \thanks{Talk presented by O. Philipsen, work supported by
        the EU under TMR programme ERBFMRX-CT97-0122. }
        and 
        H. Wittig\address{Theoretical Physics, 1~Keble Road, 
                           Oxford OX1~3NP, UK} }
       
\begin{document}

\begin{abstract}
The static potential in the 3d SU(2) Higgs model is computed by a 
variational calculation employing Wilson loops and two-meson operators.
String breaking is demonstrated numerically, the breaking scale
is determined and the results are compared with a quenched calculation.
\end{abstract}

\maketitle

\section{INTRODUCTION}

A fundamental property distinguishing full from quenched QCD is
the phenomenon of string breaking.
Wilson loop (WL) calculations 
have well established linear confinement at intermediate distances,
i.e.~the potential of two static colour charges in the 
fundamental representation bound by a gauge string 
rises linearly with their separation $r$.
For pure gauge theories this linear rise continues to infinity.
In the presence of matter fields 
the string is expected to break when its energy is large enough to
pair produce dynamical particles, which are then bound to the 
sources forming two static-light
``mesons''. The potential beyond the breaking scale 
$r_b$ then saturates at twice the meson energy.
The scale $r_b$ can be estimated by equating the 
linear potential extracted from WL's with twice the meson energy.
Although present day QCD simulations have reached such distances, 
direct evidence for string breaking and the saturation of the potential 
is not observed \cite{ukqcd}.

The problem is generic also for other theories with 
confinement. String breaking should
occur in the confinement phase of the SU(2) Higgs model \cite{evertz}
and in the adjoint potential in pure gauge 
theory, but similarly this is not apparent in 
WL calculations \cite{mic}.
This observation is paralleled in 3d gauge theories with 
and without matter fields \cite{3d}, which qualitatively 
have the same confinement and screening properties 
as in 4d.

The failure of WL calculations as well as a
recent strong coupling analysis \cite{dr} suggest
that WL's have poor overlap with the two-meson (MM) states,
and hence MM operators and their mixing with WL's should
be considered. In view of the qualitative
similarity between 3d and 4d gauge theories, it seems 
expedient to perform such an analysis
in a simpler theory, the confinement phase of the 3d SU(2) Higgs model.
Besides having bosonic fields only and being able
to simulate larger volumes, the superrenormalisability of 3d gauge theories
accounts for analytically known constant physics curves \cite{lai}.
Moreover, the mass spectrum in the confinement phase 
of the model is well known and exhibits very good 
scaling behaviour \cite{us,om}. 
Working at the same point in the confinement phase and with the notation 
as in \cite{us,om}, we approach the continuum by simulating 
three gauge couplings, $\beta_G=5.0,
7.0, 9.0$, while keeping the
continuum scalar self-coupling fixed at $\beta_G\,\beta_R/\beta_H^2 =
0.0239$ (for more details see \cite{break,us}).

\section{MIXING ANALYSIS}

\begin{figure}[hbt] 
\vspace{2.5cm}
\hspace*{-1.5cm}
\ewxy{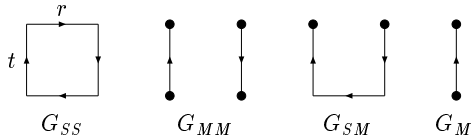}{120mm}
\vspace{-11cm}
\caption{Operators used to extract the static potential.
The dots represent scalar fields.
\label{ops}}
\end{figure}

The standard operator to compute the static potential is the 
WL of area $r\times t$, corresponding to a correlation 
$G_{SS}(t)$ of a string of length $r$, Fig.~\ref{ops}.
An operator that should be well suited
to describe the MM state after string breaking is $G_{MM}(t)$.
In order to check that this is indeed the case,
we have also measured the correlator of 
one static-light meson, given by $G_{M}(t)$.
The transition from the string to the MM state is 
represented by $G_{SM}$.

The static potential is now studied by finding
the lowest energy eigenstate contributing to the matrix correlator
at every given $r$,
\begin{equation}
 G(r,t) = \left(\begin{array}{ll}
        G_{SS}(r,t) & G_{SM}(r,t) \\
        G_{MS}(r,t) & G_{MM}(r,t)
                \end{array}\right).
\label{eq_matcor}
\end{equation}
This strategy has also been employed in calculations of the 
adjoint potential \cite{mic}.

In order to enhance the signal, we have used
smeared field variables obtained by the
standard ``fuzzing" algorithm \cite{alb} for the links and a suitable
modification of scalar smearing described in \cite{us}.
For the operator basis we selected three link and 
five scalar fuzzing levels, leaving us
with an $8\times 8$ correlation matrix 
$G_{ik}(r,t)=\langle \phi_i \phi_k \rangle$, where the $\phi_i$ stand for
string or MM operators at a given fuzzing level. 
The procedure to diagonalise
$G_{ik}$ by a variational method has been discussed in detail 
in \cite{mic,us}. 
It results in eigenvectors $\Phi_i=\sum_{k} a_{ik} \phi_k$
from which the correlation functions of the (approximate) eigenstates of the
Hamiltonian may be calculated,
\begin{equation}
  \Gamma_i(t) = \langle \Phi_i(t) \Phi_i(0) \rangle =
\sum_{j,k=1}^8\,a_{ij}a_{ik}\,G_{jk}(t).
\label{eq_eigenstate}
\end{equation}
The coefficients $a_{ik}$ are a measure for the overlap between
the operators used in the simulations and
the energy eigenstates.
This analysis not only allows to extract the ground
state energy, but also those of the first excited states
as well as the ``composition" of the mass eigenstates
in terms of the original operator basis.

\section{RESULTS}

\begin{figure}
\vspace{-2.0cm}
\ewxy{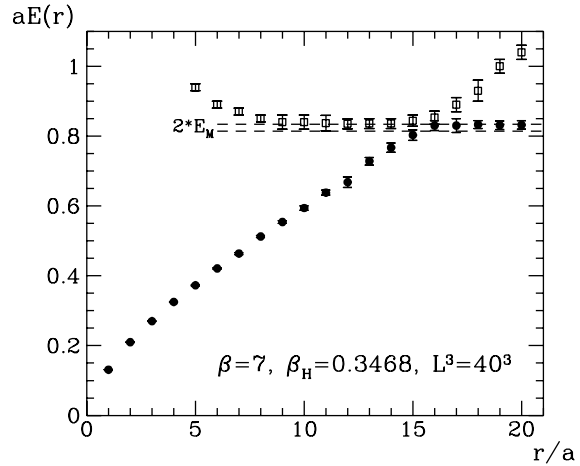}{115mm}
\vspace{-2.5cm}
\caption{The potential and its first excited state.
The dashed lines indicate
twice the energy of the single meson state, as
extracted from $G_M(t)$.
\label{pot}}
\end{figure}

In Fig.~\ref{pot} the results for the static potential and its first 
excitation are shown. 
The ground state exhibits the well known linear rise, 
but clearly levels off at the breaking 
scale $r_b$ to saturate at a constant value, which to very good
accuracy agrees with twice the static-light meson energy.
The energy of the first excited state quickly approaches the level of the
MM state for $r<r_b$, but displays a 
linear rise for $r>r_b$ in continuation of the ground state potential
below $r_b$. 

\begin{figure}
\vspace{-0.8cm}
\ewxy{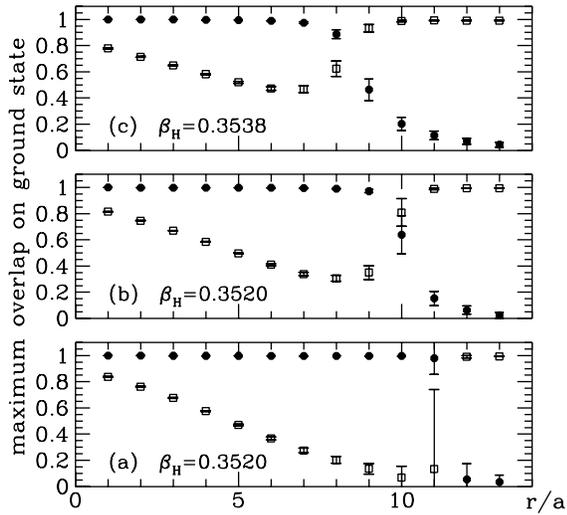}{115mm}
\vspace{-2.8cm}
\caption{The maximum projections on the ground state of the string
($a_{1,SS}$, full circles) and the MM operator
($a_{1,MM}$, open squares) for two values of the hopping parameter
and in quenched approximation (bottom)
at $\beta_G=5.0$ on $26^3$.
\label{proj}}
\end{figure}

We note that a sharp crossover in the energy levels of the string and
the MM state is also observed in a quenched calculation, which
we have performed for comparison.
In order to unambiguously establish string breaking it is necessary
to analyse the composition of the energy eigenstates as a function
of $r$ and demonstrate the decay of the string by its 
coupling to the MM state.
In Fig.~\ref{proj} the maximum projection of the string
and the MM operators
onto the ground state of the potential are shown.
For $r<r_b$, the ground state is clearly dominated by the string operator,
the MM admixture is significantly smaller, albeit non-vanishing. 
This non-vanishing
contribution survives in the quenched approximation, suggesting
two possible sources for it: First, at small distances the
two mesons may interact via gluon exchange.
Second, the smeared scalar fields
contain fuzzed links, which overlap
at high smearing levels. While the former effect is physical
the latter is a fuzzing artefact. Both 
lead to mixing with the physical string and decrease with
growing $r$.
%
In the dynamical case, there is a clear signal for mixing at $r\simeq r_b$,
with comparable
projections of both operator types onto the ground state. 
At $r>r_b$ the potential is dominated
by the MM operator,
whereas now the first excited state consists of string contributions
(not shown).
Comparing with the quenched data, we note that the
crossing in the ground state projection of the two operator types 
changes smoothly in the dynamical calculation, whereas
it is a sharp step function in the quenched case (which is  
more pronounced at $r>r_b$ where there are no fuzzing artefacts).

The properties of mixing can be studied by exploring 
different scalar masses (hopping parameters), 
c.f.~Fig.~\ref{proj}.
As expected, the string breaks earlier for lighter scalars (larger $\beta_H$). 
In qualitative confirmation of \cite{dr} we also observe
the mixing region to widen in this case, accompanied by a smoothing of the
crossover in the potential. 

Defining $r_b$ by the interpolated distance with equal projections of string
and MM operators, $a_{1,SS}(r_b)-a_{1,MM}(r_b)=0$, the continuum approach 
in Table \ref{tab_rbreak} yields
$r_{\rm b}\,g^2 \approx 8.5$, where
$m_{G}/g^2=1.60(4)$, $m_{S}/g^2=0.839(15)$ are the lightest
scalar glueball and meson states \cite{us}.
\begin{table}
\caption{
\label{tab_rbreak}
}
\begin{tabular}{cccrc}
 $\beta_G$ & $\beta_H$ & $L/a$ & \multicolumn{1}{c}{$r_{\rm b}/a$}
  & $r_{\rm b}\,g^2$ \\
\hline
 5.0 & 0.3520 & 36 &  $9.9\pm0.8$ & $7.90\pm0.64$ \\
 7.0 & 0.3468 & 40 & $14.8\pm1.0$ & $8.47\pm0.57$ \\
 9.0 & 0.3438 & 52 & $19.1\pm1.0$ & $8.50\pm0.44$ \\
\end{tabular}
\vspace{-3mm}
\end{table}

\section{CONCLUSIONS}

We have presented clear numerical evidence for string breaking in the
3d SU(2) Higgs model. Wilson loops were demonstrated not to be  
suitable to extract 
the potential for $r>r_b$. Instead, string breaking may be studied via
a mixing analysis of string and two-meson operators. 
The size of $r_b$ and the width of the mixing region
are functions of the mass of the dynamical particles. In a quenched
calculation, only the crossover in the potential, but no smooth mixing
region is observable. A similar anlaysis was also successful 
in the 4d SU(2) Higgs model \cite{ks}, and the method may be expected to
be applicable to QCD as well.

\end{document}